\documentclass[12pt]{article}
\usepackage{geometry}
\usepackage{makeidx}
\usepackage[T1]{fontenc}
\usepackage[dvipsnames,svgnames,table]{xcolor}
\usepackage{epstopdf}
\usepackage{epsfig}
\usepackage{textcomp}
\usepackage{hyperref}
\usepackage{amsmath}
\usepackage{amssymb}
\usepackage{multirow}
\usepackage{multicol}
\usepackage{booktabs}
\usepackage{lastpage}
\usepackage{hyperref} 
\usepackage{graphicx}
\usepackage{enumerate}
\usepackage{threeparttable}
\usepackage{float}
\usepackage{array}
\usepackage{cite}
\usepackage{color,soul}
\usepackage{listings}
\makeatletter
\renewcommand\section{\@startsection{section}{1}{\z@}%
                    {-2.5ex \@plus -1ex \@minus -.2ex}%
                    {2.3ex \@plus.2ex}%
                    {\normalfont\large\bfseries}}
\renewcommand\subsection{\@startsection{subsection}{1}{\z@}%
                    {-2.5ex \@plus -1ex \@minus -.2ex}%
                    {2.3ex \@plus.2ex}%
                    {\small\bfseries}}

\begin{document}
\bigskip
\title{\textbf{COME: Cylindrical oriented muon emission in GEANT4 simulations}}
\medskip
\author{\small A. Ilker Topuz$^{1}$}
\medskip
\date{\small$^1$Manipal Centre for Natural Sciences, Centre of Excellence, Manipal Academy of Higher Education, Manipal 576104, India\\ahmet.topuz@manipal.edu}
\maketitle
\begin{abstract}
In this study, a source scheme based on source biasing as well as discrete energy spectrum in the cylindrical geometry is presented for the simulations of muon tomography in the GEANT4 toolkit. First, a lateral cylindrical surface and a top circular disc act as a generation surface that surrounds the tomographic setup. Then, the generated muons are directed towards the origin where the target volume is situated. Secondly, the kinetic energy of the entering muons is assigned by using a 80-bin discrete energy spectrum between 0 and 8 GeV that is extracted from the CRY muon generator. Thus, the present recipe is called cylindrical oriented muon emission (COME). This source scheme may especially find its applications in the cases where the lateral muon detectors are utilized in order to profit from the horizontal or horizontal-like muons.
\end{abstract}
\section{Introduction}
\begin{figure}[H]
\begin{center}
\includegraphics[width=12cm]{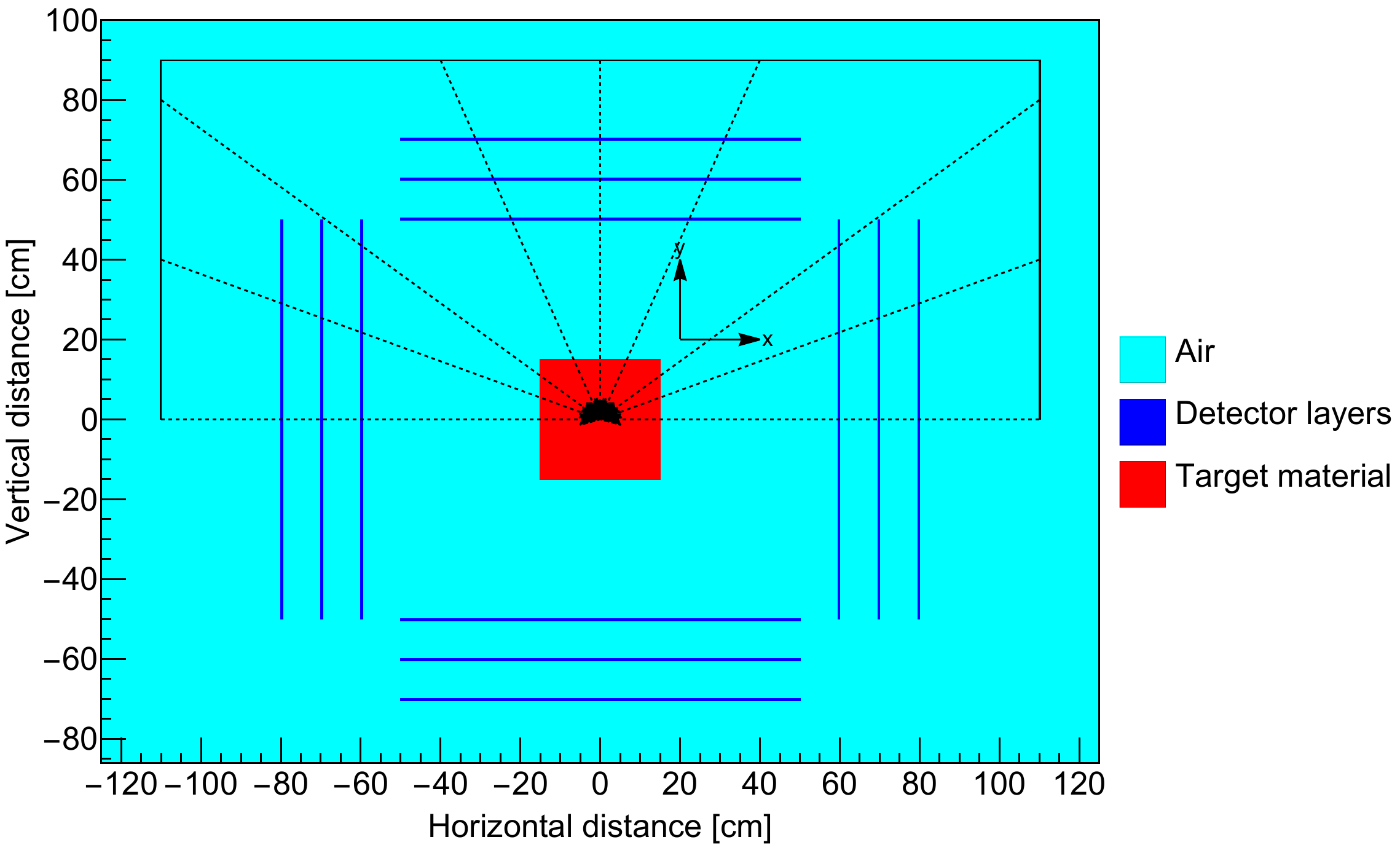}
\end{center}
\caption{Biased cylindrical source surrounding a tomographic setup with lateral detector layers.}
\label{biasedcylindricalsource}
\end{figure}

Muon tomography~\cite{bonechi2020atmospheric} is a promising technique that is practiced in order to discriminate the target volumes such as nuclear materials. Regarding the GEANT4 simulations~\cite{agostinelli2003geant4} performed in the area of muon tomography, a number of muon generators like EcoMug~\cite{pagano2021ecomug} have been already proposed by providing a couple of options in terms of the source geometry under certain conditions such as the continuous energy spectrum, the continuous angular spectrum, and the coupling between the these two spectra. By recalling the existence of the tomographic configurations where the lateral detector layers are also utilized to benefit from the horizontal muons~\cite{blackwell2015use,georgadze2021method} as shown in Fig.~\ref{biasedcylindricalsource}, the cylindrical sources might be at disposal to cover all the possible angles.

In the present study, a cylindrical source scheme hinged on the source biasing and the discrete energy spectrum is exhibited by using G4ParticleGun in the GEANT4 toolkit (see Appendix A). Both the lateral and the top surfaces are employed in order to include all the possible directions, and a fraction parameter governs the corresponding percentage of the generated muons on either surface. The generated muons on these surfaces are oriented towards the target volume via the vector construction between the generation point and the origin, and the kinetic energies are assigned through the discretization of the CRY energy spectrum between 0 and 8 GeV as described in the other studies~\cite{topuz2023particle, topuz2022dome}. This study is organized as follows. The methodology is explained in section~\ref{Methodology} where section~\ref{lateralcylindricalsurface} shows the muon emission from the lateral cylindrical surface, while section~\ref{circulartopsurface} refers to the emission from the top disc. Finally, the conclusions are drawn in section~\ref{conclusion}.
\section{Methodology}
\label{Methodology}
\subsection{Emission from lateral cylindrical surface}
\label{lateralcylindricalsurface}
The lateral surface generation is initiated by randomizing the azimuthal angle $\varphi$
\begin{equation}
\varphi=2\times \pi\times {\rm G4UniformRand()}
\end{equation}
The coordinate transformation yields the generated points on the lateral cylindrical surface for a cylinder of radius $R$ and height $h$ in the Cartesian coordinates as described in
\begin{equation}
x_{i}=R\times\cos\varphi
\end{equation}
and
\begin{equation} 
y_{i}=h\times{\rm G4UniformRand()}
\end{equation}
and 
\begin{equation} 
z_{i}=R\times\sin\varphi
\end{equation}
Then, the generated particles on the cylindrical surface are directed to the origin in order to minimize the particle loss
\begin{equation}
x_{f}=0,~~~
y_{f}=0,~~~
z_{f}=0
\end{equation}
Next, by constructing a vector from the lateral cylindrical surface to the origin, one obtains
\begin{equation}
px=x_{f}-x_{i},~~~
py=y_{f}-y_{i},~~~
pz=z_{f}-z_{i}
\end{equation}
Thus, the selective momentum direction denoted by $\vec{P}=(P_{x}, P_{y}, P_{z})$ is
\begin{equation}
P_{x}=\frac{px}{\sqrt{px^{2}+py^{2}+pz^{2}}},~~~
P_{y}=\frac{py}{\sqrt{px^{2}+py^{2}+pz^{2}}},~~~
P_{z}=\frac{pz}{\sqrt{px^{2}+py^{2}+pz^{2}}}
\end{equation}
\subsection{Emission from circular top surface}
\label{circulartopsurface}
\begin{figure}[H]
\begin{center}
\includegraphics[width=11cm]{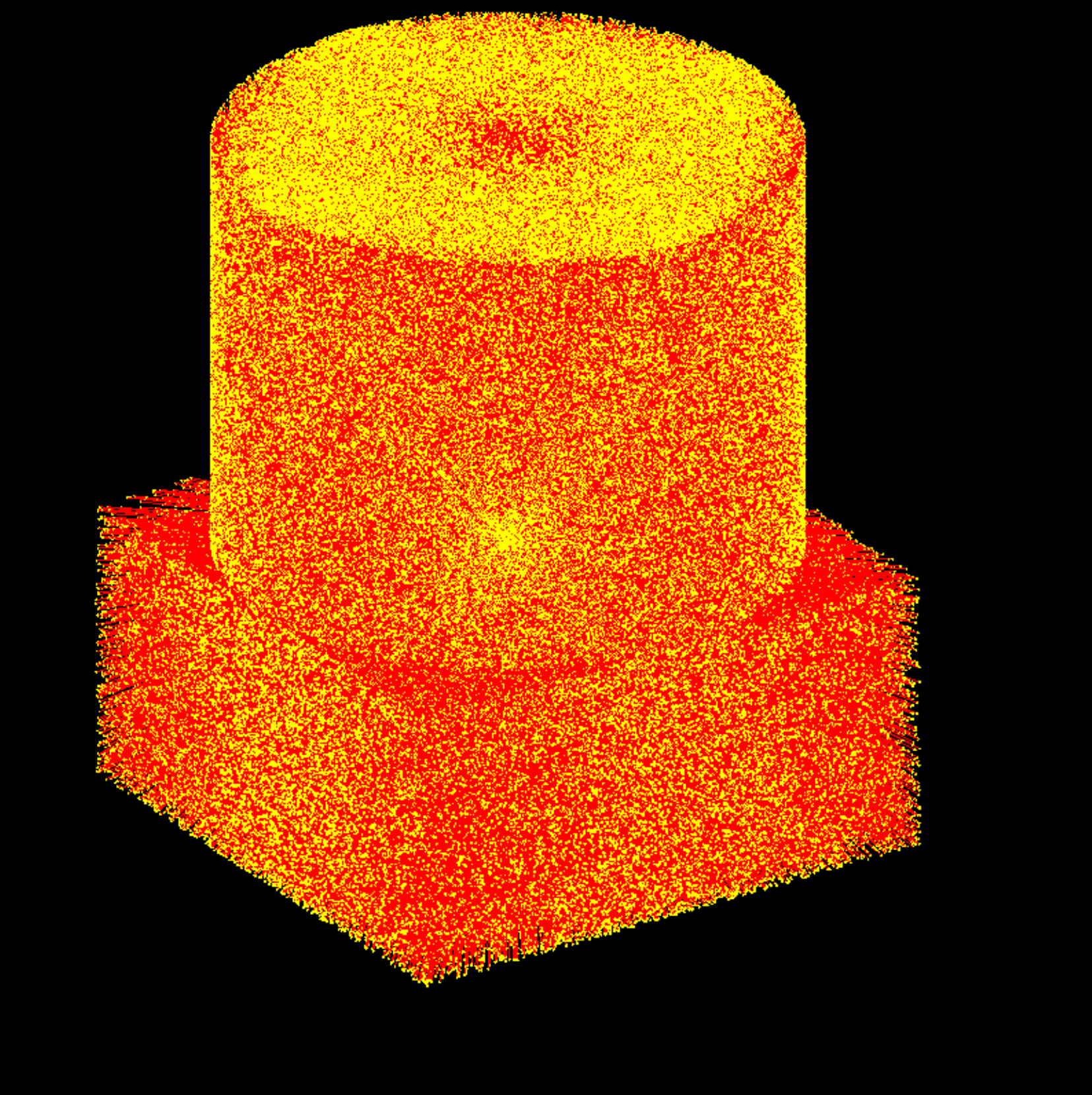}
\caption{Preview of cylindrical discrete oriented muon source with a height of 150 cm and a radius of 100 cm by using a surface fraction of 50$\%$.}
\label{preview}
\end{center}
\end{figure}

The generation of the circular top surface is also initiated by randomizing the azimuthal angle $\varphi$ as shown in~\cite{weisstein2011disk}
\begin{equation}
\varphi=2\times \pi\times {\rm G4UniformRand()}
\end{equation}
The generation points on the top disk in the Cartesian coordinates are given by 
\begin{equation}
x_{i}=R\times\sqrt{\rm G4UniformRand()}\times\cos\varphi
\end{equation}
and
\begin{equation}
y_{i}=h
\end{equation}
and
\begin{equation} 
z_{i}=R\times\sqrt{\rm G4UniformRand()}\times\sin\varphi
\end{equation}
Then, the generated particles on the circular surface are again directed to the origin
\begin{equation}
x_{f}=0,~~~
y_{f}=0,~~~
z_{f}=0
\end{equation}
Next, by constructing a vector from the disc surface to the origin, one obtains
\begin{equation}
px=x_{f}-x_{i},~~~
py=y_{f}-y_{i},~~~
pz=z_{f}-z_{i}
\end{equation}
Thus, the selective momentum direction denoted by $\vec{P}=(P_{x}, P_{y}, P_{z})$ is
\begin{equation}
P_{x}=\frac{px}{\sqrt{px^{2}+py^{2}+pz^{2}}},~~~
P_{y}=\frac{py}{\sqrt{px^{2}+py^{2}+pz^{2}}},~~~
P_{z}=\frac{pz}{\sqrt{px^{2}+py^{2}+pz^{2}}}
\end{equation}
Finally, the simulation preview through the present scheme is displayed in Fig.~\ref{preview}.
\section{Conclusion}
\label{conclusion}
To conclude, the algorithmic recipe of a biased cylindrical source scheme with a discrete energy spectrum is presented for the applications of muon tomography in the GEANT4 simulations. This scheme might be introduced into the muon generators such as CRY in order to prefer the cylindrical surfaces rather than a planar generation.
\section{Conflict of interest}
The author declares that there is no conflict of interest.
\section*{Appendix A - Biased discrete cylindrical muon source}
\begin{tiny}
\begin{lstlisting}
#include "B1PrimaryGeneratorAction.hh"
#include "G4LogicalVolumeStore.hh"
#include "G4LogicalVolume.hh"
#include "G4Box.hh"
#include "G4RunManager.hh"
#include "G4ParticleGun.hh"
#include "G4ParticleTable.hh"
#include "G4ParticleDefinition.hh"
#include "G4SystemOfUnits.hh"
#include "Randomize.hh"
#include <iostream>
using namespace std;
//....oooOO0OOooo........oooOO0OOooo........oooOO0OOooo........oooOO0OOooo......

B1PrimaryGeneratorAction::B1PrimaryGeneratorAction()
: G4VUserPrimaryGeneratorAction(),
  fParticleGun(0), 
  fEnvelopeBox(0)
{
  G4int n_particle = 1;
  fParticleGun  = new G4ParticleGun(n_particle);

  // default particle kinematic
  G4ParticleTable* particleTable = G4ParticleTable::GetParticleTable();
  G4String particleName;
  G4ParticleDefinition* particle
    = particleTable->FindParticle(particleName="mu-");
  fParticleGun->SetParticleDefinition(particle);
}

//....oooOO0OOooo........oooOO0OOooo........oooOO0OOooo........oooOO0OOooo......

B1PrimaryGeneratorAction::~B1PrimaryGeneratorAction()
{
  delete fParticleGun;
}

//....oooOO0OOooo........oooOO0OOooo........oooOO0OOooo........oooOO0OOooo......

void B1PrimaryGeneratorAction::GeneratePrimaries(G4Event* anEvent)
{
//Discrete probabilities
double A[]= {0.0, 0.01253639, 0.02574546, 0.02802035, 0.02706636, 0.03528534, 0.02826496,
0.03157946, 0.03078447, 0.02777574, 0.02546415, 0.03150608, 0.02815489,
0.02580661, 0.02364179, 0.02170935, 0.02152589, 0.02348279, 0.02134243,
0.0196913,  0.02036398, 0.01841931, 0.01718402, 0.01700056, 0.01624226,
0.01539835, 0.01536166, 0.01471344, 0.01422421, 0.01412637, 0.01284215,
0.01260977, 0.01213278, 0.0129033,  0.01248746, 0.01196155, 0.01064064,
0.01057949, 0.0096255,  0.0103838,  0.00928304, 0.00879382, 0.00884274,
0.00793767, 0.00786429, 0.00769306, 0.00709376, 0.00736283, 0.0071916,
0.00721607, 0.00692253, 0.00643331, 0.00678799, 0.00673907, 0.00618869,
0.00634769, 0.00665346, 0.00650669, 0.00561385, 0.00589516, 0.00589516,
0.00578508, 0.00557716, 0.00550378, 0.00434187, 0.0043541,  0.00408503,
0.00364472, 0.00399941, 0.00388934, 0.00396272, 0.00431741, 0.00368142,
0.00363249, 0.00362026, 0.00410949, 0.00336342, 0.00358357, 0.00362026,
0.00348573, 0.0035958}; 
//Discrete energies
double B[]= {0.0, 100, 200, 300, 400, 500, 600, 700, 800, 900, 1000, 
  1100, 1200, 1300, 1400, 1500, 1600, 1700, 1800, 1900, 2000, 
  2100, 2200, 2300, 2400, 2500, 2600, 2700, 2800, 2900, 3000, 
  3100, 3200, 3300, 3400, 3500, 3600, 3700, 3800, 3900, 4000, 
  4100, 4200, 4300, 4400, 4500, 4600, 4700, 4800, 4900, 5000, 
  5100, 5200, 5300, 5400, 5500, 5600, 5700, 5800, 5900, 6000, 
  6100, 6200, 6300, 6400, 6500, 6600, 6700, 6800, 6900, 7000, 
  7100, 7200, 7300, 7400, 7500, 7600, 7700, 7800, 7900, 8000};
G4int SizeEnergy=sizeof(B)/sizeof(B[0]);
G4int SizeProbability=sizeof(A)/sizeof(A[0]);
G4double Grid[sizeof(B)/sizeof(B[0])];
double sum=0;
  for(int x=0; x < 81; x++){
  sum=sum+A[x];
  Grid[x]=sum;
  }
  G4double radius=100*cm; //Radius of cylinder
  G4double height=150*cm; //Height of cylinder
//Centerally focused cylindrical source via coordinate transformation - by AIT
  G4double theta=2*3.14159265359*G4UniformRand();
  G4double fraction=G4UniformRand();
  G4double x0;
  G4double y0;
  G4double z0;
  if(fraction<=0.5){
//Coordinates on lateral cylindrical surface
  x0=radius*cos(theta);
  y0=height*G4UniformRand();
  z0=radius*sin(theta);
  }
//Coordinates on circular surface  
  if(fraction>0.5){
  x0=radius*sqrt(G4UniformRand())*cos(theta);
  y0=height;
  z0=radius*sqrt(G4UniformRand())*sin(theta);
  }
  fParticleGun->SetParticlePosition(G4ThreeVector(x0,y0,z0));
//Aimed at origin
  G4double x1=0;
  G4double y1=0;
  G4double z1=0;
  G4double mx = x1-x0;
  G4double my = y1-y0;
  G4double mz = z1-z0;
  G4double mn = sqrt(pow(mx,2)+pow(my,2)+pow(mz,2));
  mx = mx/mn;
  my = my/mn;
  mz = mz/mn;
  fParticleGun->SetParticleMomentumDirection(G4ThreeVector(mx,my,mz));
  G4double Energy=0; //Just for initialization
  G4double pseudo=G4UniformRand();
  for (int i=0; i < 81; i++){
  if(pseudo > Grid[i] && pseudo <= Grid[i+1]){
  Energy=B[i+1];
  std::ofstream EnergyFile;
  EnergyFile.open("Energy.txt", std::ios::app);
  EnergyFile <<  Energy << G4endl;
  EnergyFile.close();
  } 
  }   
  fParticleGun->SetParticleEnergy(Energy);
  fParticleGun->GeneratePrimaryVertex(anEvent);
}
//....oooOO0OOooo........oooOO0OOooo........oooOO0OOooo........oooOO0OOooo......
\end{lstlisting}
\end{tiny}
\bibliographystyle{ieeetr}
\bibliography{COMEbiblio} 
\end{document}